\documentstyle{EuroPhys}

\input EuroMacr

\begin{document}

\euro{}{}{}{} \Date{10 Nov 1997}

\shorttitle{ A.C. Maggs, Twist dynamics} \title{Twist dynamics of
  semiflexible polymers}

\author{A. C. Maggs}

\institute{PCT, ESPCI, 10 rue Vauquelin, 75231 Paris Cedex 05,
  France.}

\rec{}{}

\pacs{\Pacs{}{}{}}

\maketitle

\begin{abstract}
  A number of strange results have been reported for the twist
  elasticity of a semiflexible filament, actin.  In particular dynamic
  and static methods for determining the torsional twist modulus give
  very different results. I show here that internal losses (friction)
  in a semiflexible filament could be an important cause of
  dissipation and that the interpretation of dynamic experiments
  should be made in terms of a storage and loss modulus for actin
  proteins.
\end{abstract}

Several recent experiments \cite{second,twist} have tried to
characterize the torsional persistence length of actin filaments. The
most straight forward measure the static twist fluctuations by
attaching a bead to the end of a single filament and following its
torsion phase \cite{second}.  These experiments give a result which is
close to that well established for the bending elastic constant
\cite{howard} measured from transverse fluctuations; the result is to
be expected with a material with Poisson ratio close to zero
\cite{landau}. A second class of experiments \cite{twist} has always
seemed to give strange results. They depend on measuring the torsional
dynamics, and are not in agreement with the static measurements. I
have been motivated by this anomaly to reconsider the theory of
torsional dynamics of semiflexible polymers and the importance of
dynamic bend-torsion coupling.  A simple argument will show that there
can be non-hydrodynamic contributions to the dissipation and that
these could be as important as hydrodynamic drag on the twist
dynamics.  Since dynamic methods really measure the product of a
stiffness and friction coefficient the results of static and dynamic
experiments should not necessarily be the same. For this mechanism to
be important we are forced to admit that actin is described by a
viscoelastic modulus with an internal mode near 1MHz.  Since it is
well known that many proteins have slow conformational changes, this
hypothesis appears reasonable, though clearly needs experimental
confirmation for this particular protein.

Actin filaments are helical structures with a diameter, $d$ of 7 nm
and lengths which can reach tens of microns. They have a bending
persistence length which is close to 15 microns \cite{howard}.  Much
theoretical work has considered the bending dynamics of these
filaments however relatively less thought has been given to twist
dynamics in these systems; however careful theoretical work on twist
dynamics in DNA \cite{zimm,dna} has been used to interpret a series of
dynamic torsional experiments on actin \cite{twist}. The twist
dynamics can be observed by a technique of polarized fluorescent
scattering. One marks the sample with a fluorescent probe and
illuminates with a laser.  The dye absorbs a photon, which is slightly
later re-emitted.  If the dye rotates between the absorption and
reemission of the photon there is a decorrelation in the signal
compared with a stationary object.

The rest of this article is organized as follows. We first summarize
the rotational dynamics of semiflexible polymers in the limit of
linear dynamics.  We then note that the experimental situation in
actin is expected to be out of the range of validity of the simple
linear theory, in particular the {\it dissipation} links the twist and
bend degrees of freedom. This coupling can lead to extra internal
losses which we shall estimate in a rather crude manner.

In linear theory there are two main contributions to the rotation of
the dye fixed on a filament.  Firstly the internal modes of the
filament obey a Rouse like equation
\begin{equation}
 \rho_{\phi}{{\partial \phi} \over {\partial t}} = K { {\partial^2 \phi} \over {
    \partial s^2}} +f(t,s) \label{twist}
\end{equation}
where $\phi(s)$ is the angle turned by a filament with respect to the
fluid, $K$ is the twist stiffness and $\rho_{\phi}$ a friction
coefficient. $f(s,t)$ describes the Brownian noise of the fluid.
Secondly the bending of the filaments in space leads to a rotation of
the tangent of the filament. The bending dynamics of a filament are
described by a Langevin equation \cite{hearst,farge}
\begin{equation}
\rho {{\partial {\bf r}_\perp} \over {\partial t}}= -\kappa { {\partial ^4
    {\bf r_{\perp}} \over { \partial s^4 }}} +{\bf f_{\perp}}(s,t)
\label{bend}
\end{equation}
With $\kappa$ the bending elastic constant, $\rho$ a dissipation
coefficient and ${\bf r_{\perp}}$ the transverse amplitude of the
fluctuations of the filament. The direction of the filament is given
by its tangent, $ {\bf t} = \partial_s {\bf r(s)}$.

The rotation of a monomer on a filament (measured in a dynamic
experiment) has two contributions
\begin{equation}
<(\phi(s,t) - \phi(s,0))^2 >\sim \int {{1-\exp(-2 K t q^2/\rho_{\phi})} \over
  {Kq^2}} dq \sim  (t/\rho_{\phi} K)^{1/2}
\label{twist2}
\end{equation}
from eq. (\ref{twist}) and
\begin{equation}
\label{bend2}
<( \partial_s {\bf r_{\perp}} (t) -  \partial_s {\bf r_{\perp}}    (0))^2>  \sim 
\int q^2 { {1- \exp(-2 \kappa q^4 t/\rho)} \over {\kappa q^4} } dq 
 \sim   (t/\rho \kappa^3)^{1/4}
\end{equation}
from eq. (\ref{bend}).  If we look at the time scales in the eq.
(\ref{twist}) and (\ref{bend}) we see that there are two relaxation
times $\tau_{t}$, the twist relaxation time and $\tau_{b}$ the bend
relaxation time which scale differently with the excitation
wavelength, $q$. In fact $\tau_b/\tau_t\sim(qd)^2$ with $d$ the
diameter of the filament.  Experimentally for a ten micron wavelength
mode we expect that $\tau_b$ should be several seconds while $\tau_t$
is rather $100 \mu s$. Given the great disparity in time scales one
expects that (\ref{twist2}) is the dominant term in the twist dynamics
in most experimental situations (though in gels both should be
important since the free rotation of the filaments is impossible).
Dynamic depolarization experiments are then sensitive to correlation
functions of the form
\begin{equation}
 I(t) = \exp ( - \alpha (\Phi(s,t) - \Phi(s,0))^2) 
\label{sum}
\end{equation}
Where $\Phi$ is a sum of the two contributions (3) and (4) leading to
a stretched exponential decay of the signal. For more detail see
\cite{zimm}, $\alpha$ is of order unity.

These two independent contributions are derived, however, via
linearisation of the dynamic equations.  We shall now make the
argument that for long filaments the twist and bending are strongly
coupled and that this leads to important modifications of eq.
(\ref{twist}) and (\ref{bend}): When filaments are very short the
simple twist and bend modes are eigenvalues of both the static and the
dynamic problem, since the friction matrix is diagonal and does not
couple the bend and twist degrees of freedom. For longer filaments the
situation is more complicated.  There is no coupling in the energy but
there is a dynamic coupling in the friction. As an illustrative
example consider a crankshaft shaped filament. If the filament is
twisted at its ends in a resistive fluid a bending force is exerted on the
filament. Thus we are force to reconsider the nature of the dynamic
eigenmodes for long filaments.

For a long filament one can envisage two extreme situations. Either
(a) that the filament keeps its shape in body frame and thus changes
its shape rapidly in the laboratory frame i.e. that it does spin as a
crankshaft. Or (b) The filament keeps a nearly constant shape in the
laboratory frame and spins rapidly in a narrow tube, implying that the
filament undergoes rapid bending in the body frame.  Experimentally
one sees under a microscope no evidence of case (a), one also
understands that rapid rotation of a constant shaped filament will be
unfavorable due to the high friction coefficient of the spinning
object.  We conclude that superposed on the slow shape fluctuations of
a actin filament is a very rapid spinning of the filament which forces
the filament to bend at high frequencies, typically 10 $kHz$.
More formally we know that a general non-planar curve is characterized
by non-zero torsion and writhe. Even if we suppress completely the
torsional fluctuations (described by the variable $\phi$) geometry
implies that bending fluctuations necessarily lead to rotational
movement of the filament and thus additional contributions to the
dissipation. 

We now argue that because of the large ratio of the relaxation times
of the bending and twisting modes we can use simple physical arguments
to deduce the appropriate dynamical variables in this coupled,
non-linear problem. We first consider the case of a filament embedded
in a gel (or almost equivalently in a dense solution of filaments).
In this case the long wavelength bending degrees of freedom are
completely frozen out.  However the filament can continue to twist
freely according to eq.  \cite {twist} if we interpret $\phi$ as
rotation in the tube of the gel. If we now relax the absolute
constraint on the bending degrees of freedom imposed by a gel we
realize that the large ratio of $\tau_b$ to $\tau_t$ leads to an
almost identical situation; the bending modes can not relax on a time
scale of the twist motion. We can deduce that the two dynamic modes of
the filament in free solution are (1) A rapid twist mode in a narrow
tube which is described by eq. (\ref{twist}), which however leads to
strong bending fluctuations in the body frame; and (2), a slow bend
mode which is followed adiabatically by the much faster twist degrees
of freedom.  For frequencies lower than a few $kHz$ this mode should be well
described by eq. (\ref{bend}).  We now pass on to the consequences of
these remarks on the calculation of the twist dynamics.

The energetics of semiflexible polymers are usually modeled using a
description in terms of a simple elastic theory of a uniform rod
\cite{landau} or more accurately a helical structure \cite{siggia}. In
these models the filament is taken to perfectly elastic; that is the
storage modulus is independent of frequency. In general we expect that
at high frequencies deformations are going to be couple to internal,
dissipative modes and that a more accurate description is in terms of
a viscoelastic, Voigt, solid \cite{ferry}, described by the
constitutive equation
\begin{equation} E(\omega) = { {E_0} \over {1+i \omega \tau} } \label{voigt} \end{equation}
where $E_0$ is the low frequency elastic modulus (measured in static
experiments) and $\tau$ is a characteristic internal time scale.
Experiments performed at finite frequency are thus coupled to both the
real and imaginary part of the response function.  We shall now show
that a simple estimate of the dissipation coming from eq.(\ref{voigt})
leads to a modification of the friction coefficient in (\ref{twist})
if the internal time scale in the constitutive equation is slower than
$10^{-6}s$.  We not that the case of a broad spectrum of modes should
be even more interesting: In particular in the case of a material with
many low frequency modes the dynamics should become highly anomalous.

The material equation eq. (\ref{voigt}) can be written in terms of
storage and loss moduli. At frequencies satisfying $\omega \tau <<1$
eq. (\ref{voigt}) has the following interpretation: The maximum stored
energy of an excitation of amplitude $x$ varies as $e = E_0 x^2/2$.
However, there is also a weak dissipation of energy equal to
\begin{equation} dE_1/dt =  e \omega^2 \tau \label{e1} \end{equation}
We can also write the conventional hydrodynamic dissipation in a
similar form, so that if we spin a rod of diameter $d$, length $l$ at
angular frequency $\omega$ in a fluid of viscosity $\eta$ the
dissipation, is equal to
\begin{equation}
 dE_2/dt = \pi \omega^2 d^2 l \eta 
\label{e2}
\end{equation}
The fact that both $\dot E_1$ and $\dot E_2$ are quadratic in $\omega$
shows that we have a renormalization of the rotational frictional
coefficient for bent filaments.
The most delicate part of our argument now follows where we shall try
to estimate $e$ for a filament which has been bent by thermal
fluctuations.

Very short filaments are essentially straight, they can spin without
hindrance along their long axis and there is no coupling between the
bend and twist modes.  Longer filaments are described by transverse
fluctuations which scale as \begin{equation} l_{\perp}^2 \sim
  l_{}^3/l_p \label{threehalfs}
\end{equation}
with $l_{\perp}$ the amplitude of fluctuations, $ l_{} $ the length of
filament and $l_p$ the persistence length \cite{odjik}. When these
transverse fluctuations have an amplitude comparable to the diameter
of the filament the rotational friction of filament will increase
rapidly with the filament length. For filaments longer than this, as
we have argued, it is much easier for the filament to bend while
spinning and the twist eigenmode in the laboratory frame is in fact a
combined twist-bend mode.  Thus the coupling of the bend to the twist
will become important when $l \sim (d^2 l_p)^{1/3}$. If we make a
filament of this length oscillate an amplitude $d$ we are in turn
stocking an energy comparable to $k_BT$ in each section of length $l$.
Thus we argue that for $e$ we should count $k_BT/(d^2 l_p)^{1/3}$ per
unit length of filament.  We are now in a position to compare the two
sources of dissipation: The ratio of the two dissipation rates varies
as
\begin{equation}
  R \sim {{\omega^2 d^2 (d^2 l_p) ^{1/3} \eta } \over
    {k_BT \omega^2 \tau}}
\label{ratio}
\end{equation}

These two sources of dissipation become comparable if we let $\tau$ be
equal to $10^{-6}s$. As noted above this does not appear too
unreasonable for a protein, however independent verification of this
point would be useful.

A particularly interesting remark in \cite{twist} implies that the
twist dynamics are only weakly dependent on the viscosity of the
solvent when this is changed by adding glucose. This would seem to be
strong evidence that the dissipation is dominated by non-hydrodynamic
sources. However, the values of the twist stiffness which are deduced
dynamically are smaller than those deduced in static experiments. From
the scaling in eq. (3) of a characteristic time in the product
$\rho_{\phi} K$ one would rather expect the dynamic measurement to produce
larger values than the static experiments. The numerical fitting
procedure used in \cite{twist} is, however, particularly complicated
including corrections due to short filaments which make a simple
re-fitting of the data essentially impossible; a full re-analysis of
the data should perhaps be done with a renormalized value of the
effective rotational friction coefficient.

We note that for length scales smaller than $l$ we should not be
coupling the twist and bend modes. Thus one can expect some sign of a
dynamic crossover for filaments or mode of scale $l$, or at
frequencies comparable to $k_B T l_p/l^4 \eta$ in the bending
dynamics. This is probably at some tens of $kHz$, frequencies which
are accessible through DLS or microrheology.
\cite{schmidt,fred,wirtz}. It would be interesting to try and make a
more quantitative theory of bend twist coupling by using the large
ratio of time scales as an expansion parameter in a adiabatic
expansion of the dynamics in order to produce a quantitative theory of
this crossover.

We note that the scale $l$ by numerical coincidence is comparable to
the helical pitch of actin filaments. One expects that the two
principal bending elastic constants of a filament are not quite the
same due to the low symmetry \cite{siggia}. If the filament is forced
to spin in a constant shape this could introduce further dynamic
couplings which should be investigated.  In the literature there are
also many reports of the presence of relatively rare ``defects'' or
``kinks'' in actin filaments so that the equilibrium form of a
filament consists of long straight segments followed by turns by
several degrees. Such kinks, if they are present, must have a major
effect on twist dynamics and can lead to further novel contributions
to the relaxation function. The presence of such kinks however would
considerably complicate the fitting of the experimental data.

We should contrast our discussion here with a recent article
\cite{kamien} in which the twist dynamics are hypothesized as being
ultra slow due a local conservation law for the writhe.  This ultra
slow twist relaxation in turn couples to the bending modes due to the
geometry of twist leading to slow relaxation of bending modes. We see
here that the torsional modes are very rapid compared to the bending
modes.  This is due to the fact that the relaxation time of the
bending modes scales as $q^4$, while the twist modes scale more slowly
in $q^2$.  It seems unlikely that such considerations of ultra slow
dynamics apply in experiments on actin solutions.

\stars 

\end{document}